*Phylogenetics*

# GeneSupport – Maximum Gene-Support Tree Approach to Species Phylogeny Inference

Yunfeng Shan[1,*] and Xiu-Qing Li[2]

[1]School of Computer Science, University of Windsor, 401 Sunset Avenue, Windsor, ON N9B 3P4

[2]Molecular Genetics Laboratory, Potato Research Centre, Agriculture and Agri-Food Canada, 850 Lincoln Rd, P.O. Box 20280, Fredericton, New Brunswick, E3B 4Z7, Canada.

**ABSTRACT**

**Summary:** GeneSupport implements a genome-scale algorithm: Maximum Gene-Support Tree to estimate species tree from gene trees based on multilocus sequences. It provides a new option for multiple genes to infer species tree. It is incorporated into popular phylogentic program: PHYLIP package with the same usage and user interface. It is suitable for phylogenetic methods such as maximum parsimony, maximum likelihood, Baysian and neighbour-joining, which is used to reconstruct single gene trees firstly with a variety of phylogenetic inference programs.

## 1 INTRODUCTION

Sequences from genes, proteins, and genomes are increasing rapidly with progress of genome projects. Reconstruction of phylogenies started to use large data sets involving hundreds of genes. This paper introduces a computer program: GeneSupport. It is a tool for estimating species tree from gene trees through comparing many gene trees and computing gene supports of unique gene trees. The GeneSupport program is implemented in C based on the maximum gene-support tree approach proposed by Shan and Li (2008), which described an alternative approach to evaluate the reliability of species phylogeny inferences based on gene trees (Shan and Li, 2008). It also describes a biologic phenomenon (intuitively obvious): that closely related species share similarities in a higher number of orthologous genes than distantly related species. It is mentioned that "I am very sure that many people working on phylogenetics will find it helpful, as an alternative to the mere concatenation of separate gene sequences" (from an anonymous reviewer in 2008).

## 2 METHODS

**Computational steps:** Distances between tree pairs are computed. Two trees with identical topology have a tree distance of zero. The number of unique trees with identical topology are counted. Distances are computed based on the widely known Symmetric Difference of Robinson and Foulds (1981). The Symmetric Difference ignores branch length information, only use the tree topologies. This is the minimum number of steps required to convert between two trees, that is, the number of branches that differ between a pair of trees (Robinson and Foulds, 1981). The Robinson and Foulds topological distance is an important and frequently used tool to compare phylogenetic tree structures (Makarenkov and Leclerc, 1999; 2000). It is widely used in PHYLIP (Felsenstein, 1989) or PAUP* (Swofford, 2002) packages. We used some code from program: treedist of PHYLIP package, especially the function for computing the symmetric distances between trees because the function is well tested extensively, which is used with the kind permission of Dr. Joseph Felsenstein. Although the examples we have discussed have involved fully bifurcating trees, the input trees can have multifurcations. For the Symmetric Difference, it can lead to distances that are odd numbers.

**Restriction:** However, one strong restriction must be noted. The trees should all have the same list of species. If you use one set of species in the first two trees, and another in the second two, the distances will be incorrect and will depend on the order of these pairs in the input tree file, in odd ways.

**Gene-support and maximum gene-support tree:** The index of gene-support is the number of genes that infer a unique topology, which is equal to the tree frequency when single gene trees are inferred from single genes. The numbers of genes were calculated for all unique gene trees from the phylogenetic reconstruction results. A maximum gene-support tree was defined as a unique tree that was inferred by the highest number of genes among all the gene trees generated. Users may infer gene trees firstly with popular phylogentic methods such as maximum parsimony, maximum likeligood (ML), Baysian, neighbour-joining (NJ) seperately by means of ad hoc phylogenetic analysis packages such as PHYLIP, PAUP*, Mr.Bayes, PAML and so on.

The usage and interface of GeneSupport are similar to those of PHYLIP package for users'convenience for those who experience in the popular PHYLIP package.

---

[*]To whom correspondence should be addressed.

## 3 DEMONSTRATION

The GeneSupport analysis is demonstrated on a sample of tetrapod origin study for 43 genes from 7, 6 taxon sets (Supplementary Materials). As shown in Table 1, maximum gene support tree approach clearly showed that gene supports for four types of trees were not evident different, so 43 genes were not able to resolve the phylogenetic relationship for these 7-taxon set whatever the phylogentic methods were used. It is recognized that 43 genes did not reach the minimum requirement of the genes for inferring species tree from gene trees for the 7-taxon set.

Currently, number of sampled genes seems to be arbitrary. When a reliable tree is not known, determination of minimum required genes is difficult. 100% bootstrap support does not mean that the branch is 100% correct. 100% bootstrap support may occur in an alternative branch (Phillips et al., 2004) High bootstrap support does not necessarily signify 'the truth' (Soltis et al. 2004). When a maximum gene-support value is not evidently different, for example, in the case of 7 taxa, it can be recognized that the number of genes used does not meet the requirement of minimum genes. More genes or less taxa is necessary to be re-sampled. This is the outstanding advantage of the maximum gene support tree approach.

**Table 1**. Gene supports for four tree types of 7 taxa inferred with three methods

| Methods | Type of Trees | | | |
|---|---|---|---|---|
| | Tree I | Tree II | Tree III | Tree IV |
| MP | 2 | 2 | 2 | 2 |
| ML | 2 | 1 | 2 | 0 |
| NJ | 2 | 1 | 1 | 0 |

Table 2 showed that when taxon set reduced to 6 taxa, tree II (Fig 1 in Supplementary Materials) was supported by much less genes than tree I or tree III inferred with maximum parsimony (MP), which significant differences were observed for MBACLR and MACLRS ( Table 2) at p < 0.10 level by means of chi-square test. There were no significant differences in the gene supports of tree I and tree III. Tree IV was supported by one gene for the taxon set of MBCLRS only (Table 2). So, we rejected tree II and tree IV based on significant low gene supports. However, we still could not determine which is the maximum gene-support tree as species tree from tree I and tree III in this case. More genes or less taxa is necessary to be re-sampled. Other three demonstrations were performed for yeasts, plants and microorganisms and their maximum gene-support trees as species trees were successfully identified (Shan and Li, 2008).

**Table 2**. Gene supports for four tree types of 6 taxa inferred with MP

| Taxon Set | Type of Trees | | | |
|---|---|---|---|---|
| | Tree I | Tree II | Tree III | Tree IV |
| MBACLR | 6 | 1[+] | 6 | 0 |
| MBACLS | 4 | 3 | 5 | 0 |
| MBCLRS | 5 | 3 | 7 | 1 |
| MACLRS | 1 | 1[+] | 6 | 0 |
| BACLRS | 2 | 3 | 3 | 0 |

Notes: The 7 taxa included: Mammal (M), Bird (B), Amphibian (A), Coelacanth (C),
Lungfish (L), Ray-finned Fish (R), and Shark (S). [+, *] indicated chi-square test significant level α at P <0.10, 0.05 between the frequencies of tree II and tree I/III, respectively.

## ACKNOWLEDGEMENTS

We thank Dr. J. Felsenstein for permit the use of his publicly available code from the program treedist of PHYLIP package.